\documentclass[12pt]{elsarticle}

\usepackage{amssymb}
\usepackage{tensor}

\usepackage{amsmath}
\usepackage{xcolor}
\usepackage{subcaption}
\usepackage{dsfont}

\journal{arXiv}

\usepackage{hyperref}
\hypersetup{
     colorlinks=true,
      linkcolor=blue,
}

\begin{document}

\begin{frontmatter}



\title{Equations of motion for charged particles in strong laser fields}


\author{H. Ruhl and C. Herzing}

\address{Ludwig-Maximilians-University Munich, Theresienstrasse 37,
  80333 Munich, Germany}

\begin{abstract}
Starting from the Dirac equation coupled to a classical radiation
field a set of equations of motion for charged quasi-particles in the
classical limit for slowly varying radiation and matter fields is
derived. The radiation reaction term derived in the paper is the
Abraham-Lorentz-Dirac term. 
\end{abstract}


\end{frontmatter}



\section{Introduction}
\label{sect_introduction}
The interaction of electrons and positrons with their radiation field
is described by the Dirac equation coupled to Maxwell's equations.

The goal of the present paper is to outline the derivation of a
dynamical framework for charged quasi-particles in the classical limit
neglecting spin for slowly varying matter and radiation fields from
first principles.

The present paper is structured as follows: First, a classical Vlasov
equation is derived for spinless electrons and positrons coupled to
Maxwell's equations from the fundamental theory
of electromagnetism. Next, the concept of quasi-particles for scalar
electrons and positrons is introduced. In a third step dynamical
equations for the energy-momentum tensors of the matter and radiation 
fields are derived. The latter are utilized to obtain a set of
classical molecular dynamical (MD) equations of motion for electrons
and positrons coupled to their retarded radiation fields. 
The Lorentz-Abraham-Dirac (LAD) term for radiation reaction is 
obtained. The LAD term, however, is not the only radiation reaction
force that can be derived with the help of the methodology presented 
in this paper.

\section{Matter and radiation fields}
\label{framework}
We start by defining the concept of a Wigner operator outlined in
\cite{VASAK1987462}. The Wigner operator is
\begin{eqnarray}
\label{gauge_covariant_wigner_operator}
&&\tensor{\hat{W}}{_d_b} \left( x, p \right) =\int \frac{d^4y}{(2\pi \hbar)^4} \, e^{-i \frac{p \cdot y}{\hbar}} \,
\tensor{\hat{\Psi}}{_{d} _{b}} \left( x+\frac{y}{2}, x-\frac{y}{2}
   \right) 
\end{eqnarray}
with the kernel
\begin{eqnarray}
&&\tensor{\hat{\Psi}}{_{d} _{b}} \left( x_1, x_2 \right) = \tensor{\bar{\psi}}{_b} \left( x_1 \right) \, U 
   \left( A, x_1, x_2 \right) \, \tensor{\psi}{_d} \left( x_2 \right)
   \, ,
\end{eqnarray}
where
\begin{eqnarray}
&&U \left( A, x_1, x_2 \right) \\
&&=\exp \left[ -\frac{ie}{\hbar} \, \left( x_1 -x_2 \right)^{\nu} \int^1_0 ds \, A_{\nu}
   \left( \frac{x_1 + x_2}{2} +\left\{  s - \frac{1}{2} \right\} \left(
   x_1 -x_2 \right)  \right) \right] \, . \nonumber 
\end{eqnarray}
With the help of the Dirac equation coupled to
the radiation field equations of motion for the Wigner operator
(\ref{gauge_covariant_wigner_operator}) are
obtained \cite{VASAK1987462}. In the limit of a slowly varying
classical radiation field we obtain
\begin{eqnarray}
\label{eqm:gauge_covariant_wigner_talk1}
&&\left[ m \, \tensor{ \left( \mathds{1} \right)}{_c _a} -
   \tensor{\left( \gamma^{\mu} \right)}{_c _a} \,
   \left( p_{\mu} + \frac{i}{2} \left( \partial^x_{\mu} -e \,
   \tensor{F}{_{\mu} _{\nu}} \left( x \right) \partial^{\nu}_p \right) \right) 
   \right] \, \tensor{\hat{W}}{_a _b} \left( x, p \right) 
   =0 \, .
\end{eqnarray}
Next, it is useful to expand the Wigner operator in spin
space \cite{VASAK1987462}. This yields
\begin{eqnarray}
\label{spin-space-expansion-wigner}
&&\hat{W}_\mathrm{ab} \left( x, p \right) = \left[ \left( 1 + \tensor{\gamma}{^\beta}
   \frac{\tensor{p}{_\beta}}{m} \right) \, \left(
   \hat{\cal{F}} + \tensor{\gamma}{^\alpha} \gamma_5
   \tensor{\hat{\cal{A}}}{_\alpha} \right) \right]_\mathrm{ab} \left(
   x, p \right) \, ,
\end{eqnarray}
where $\hat{\cal{F}}$ is a scalar and
$\tensor{\hat{\cal{A}}}{^\alpha}$ is an axial vector. It is found that
in the classical limit
\begin{eqnarray}
\label{material_eqnmotion_classical_scalar}
&&\tensor{p}{_{\mu}} \left( \tensor{\partial}{^\mu}_\mathrm{x} - e \,
   \tensor{F}{^\mu_\nu} \left( x \right) \, \tensor{\partial}{^\nu}_\mathrm{p} \right) 
   \, \hat{\cal{F}} \left( x, p \right) =0 \, ,  \\
\label{material_eqnmotion_classical_axial}
&&\tensor{p}{_{\mu}} \left( \tensor{\partial}{^\mu}_\mathrm{x} - e \,
   \tensor{F}{^\mu_\nu} \left( x \right) \, \tensor{\partial}{^\nu}_\mathrm{p} \right) 
   \, \tensor{\hat{\cal{A}}}{^\alpha} \left( x, p \right) =e \, \tensor{F}{^\alpha^\nu}
   \left( x \right) \, \tensor{\hat{\cal{A}}}{_\nu} \left( x, p
   \right)
\end{eqnarray}
hold. Both equations come along with the following constraints
\begin{eqnarray}
\label{material_eqnconstraint_classical_scalar}
&&\left( p^2 - m^2 \right) \, \hat{\cal{F}} \left( x, p \right) =0 \,
   , \\
\label{material_eqnconstraint_classical_axial}
&&\left( p^2 - m^2 \right) \, \tensor{\hat{\cal{A}}}{^\mu} \left( x, p
   \right) =0 \, , \quad \tensor{p}{_\mu} \,
   \tensor{\hat{\cal{A}}}{^\mu} \left( x, p \right) =0 \, . 
\end{eqnarray}
The $4$-current is given by
\begin{eqnarray}
\label{4-current-spinor}
&&\tensor{\hat{j}}{^\mu}^\mathrm{M} \left( x \right) =-e \, \int d^4p
   \, \tensor{\gamma}{^\mu}_\mathrm{ab} \, :
   \tensor{\hat{W}}{_b_a} \left( x, p \right) : \\
&&\hphantom{\tensor{\hat{j}}{^\mu}^\mathrm{M} \left( x \right)}=-e \, \int d^4p \,
   \frac{\tensor{p}{^\mu}}{m} \, :\hat{\cal{F}} \left( x, p \right) : \nonumber
\end{eqnarray}
and the energy-momentum tensor by
\begin{eqnarray}
\label{energy-momentum-spinor}
&&\tensor{\hat{t}}{^\mu^\nu}^\mathrm{M} \left( x \right)  = \int d^4p
   \; \tensor{p}{^\nu} \tensor{\gamma}{^\mu}_\mathrm{ab} 
   \, : \tensor{\hat{W}}{_b_a} \left( x, p \right) : \\
&&\hphantom{\tensor{\hat{t}}{^\mu^\nu}^\mathrm{M} \left( x \right)}= \int d^4p \,
   \frac{\tensor{p}{^\mu}\tensor{p}{^\nu}}{m} \, :\hat{\cal{F}} \left(
   x, p \right) : \, , \nonumber
\end{eqnarray}
where normal ordering is implied. In what follows we neglect spin. The
associated classical radiation field is obtained with the help of the
ensemble-averaged current
\begin{eqnarray}
\label{maxwell-constraint}
&&\tensor{\partial}{_\mu} \tensor{F}{^\mu^\nu} \left( x \right) =\frac{1}{\epsilon_0 
   c^2} \, \tensor{j}{^\nu}^\mathrm{M} \left( x \right) \, , \quad
   \tensor{j}{^\nu}^\mathrm{M} \left( x \right) =\langle
   \tensor{\hat{j}}{^\nu}^\mathrm{M} \left( x \right) \rangle \, .
\end{eqnarray}
In addition to the energy-momentum tensor of the Dirac field
(\ref{energy-momentum-spinor}) the radiative energy-momentum tensor is
needed. It is given by
\begin{eqnarray}
\label{energy-momentum-tensor-radiation}
&&\tensor{t}{^{\mu}^{\nu}}^R \left( x \right) = \epsilon_0 c^2 \, \left[
   \tensor{F}{^{\mu}_{\alpha}} \, \tensor{F}{^{\alpha}^{\nu}} 
+\frac{\tensor{g}{^{\mu}^{\nu}}}{4} \, \tensor{F}{_{\alpha}_{\beta}}
   \, \tensor{F}{^{\alpha}^{\beta}} \right] \left( x \right) \, ,
\end{eqnarray}
where 
\begin{eqnarray}
&&\tensor{F}{^\mu^\nu} = \left(
\begin{array}{cccc}
0 & -\frac{E_x}{c} & -\frac{E_y}{c} & -\frac{E_z}{c} \\ 
\frac{E_x}{c} & 0 & -B_z & B_y \\ 
\frac{E_y}{c} & B_z & 0 & -B_x \\ 
\frac{E_z}{c} & -B_y & B_x & 0 
\end{array}
\right) \, . 
\end{eqnarray}

\section{On-shell scalar Vlasov equation}
Next, we decompose $\hat{\cal{F}}$ into positive and negative energy parts  
due to the contraint equation 
(\ref{material_eqnconstraint_classical_scalar}). We find 
\begin{eqnarray}
\label{wigner-energy-expansion}
&&\hat{\cal{F}} \left( x, p \right) =m \, \delta \left( p_0 -
   \frac{E_p}{c}\right) \, \hat{\cal{F}}^+\left( x, \vec{p} \right) \\
&&\hphantom{\hat{\cal{F}} \left( x, p \right)=}
   + m \, \delta \left( p_0 + \frac{E_p}{c}\right) \, \hat{\cal{F}}^-
   \left( x, \vec{p} \right) \, . \nonumber
\end{eqnarray}
An on shell equations is obtained by performing energy averaging. We
obtain 
\begin{eqnarray}
\label{vlasov-three-notation}
&& \left( \partial_{t} + \vec v \cdot \partial_{\vec x} \right) \left[
   f \left( x, \vec p \, \right) + \bar{f} \left( x, \vec p \, \right)
   \right] \\
&& -e \left( \vec E + \vec{v} \times \vec B \right) \cdot \partial_{\vec p} 
\left[  f \left( x, \vec p \, \right) - \bar{f} \left( x, \vec p \,
   \right) \right] =0 \, , \nonumber 
\end{eqnarray}
where following the outline in \cite{ZHUANG1996311}
\begin{eqnarray}
\label{onshell-particle}
&& f \left( x, \vec p \right) = +\frac{E_p}{c} \, \langle \hat{\cal{F}}^+ \left( x,
   +\vec{p} \, \right) \rangle \, , \\
\label{onshell-antiparticle}
&& \bar{f} \left( x, \vec p \right) = -\frac{E_p}{c} \, \langle \hat{\cal{F}}^- \left( x,
   -\vec{p} \, \right) \rangle \, .
\end{eqnarray}
Equation (\ref{vlasov-three-notation}) is the desired scalar Vlasov
equation for particle and anti-particle distributions $ f \left( x,
  \vec p \, \right)$ and  $\bar{f} \left( x, \vec p \, \right)$. 
Finally (\ref{vlasov-three-notation}) has to be augumented with
Maxwells's equations given by
\begin{eqnarray}
\label{maxwell-constraint-new}
&&\tensor{\partial}{_\mu} \tensor{F}{^\mu^\nu} \left( x \right) = -\frac{e}{\epsilon_0 
   c^2} \, \int \frac{d^3p}{p^0} \, c 
   \tensor{p}{^{\nu}} \, \left[ f \left( x, \vec{p} \right)- \bar{f}
   \left( x, \vec{p} \right) \right] \, .
\end{eqnarray}

\section{The concept of quasi-particles}
We depart from $f$ and $\bar{f}$ defined as continuous functions
on phase space and make the ansatz 
\begin{eqnarray}
\label{quasi-part}
&&g\left( \vec x, \vec p, t \right)=\sum_q \delta^3 \left( \vec x - \vec x_q (t) \right) \, \delta^3 
   \left( \vec p - \vec p_q(t) \right) \, , \\
\label{quasi-antipart}
&&\bar{g}\left( \vec x, \vec p, t \right) =\sum_q \delta^3 \left( \vec x - \vec{\bar{x}}_q (t) \right) \, \delta^3 
   \left( \vec p - \vec{\bar{p}}_q(t) \right)
\end{eqnarray}
to approximate the on shell distribution functions (\ref{onshell-particle}) and
(\ref{onshell-antiparticle}). We require that 
\begin{eqnarray}
\label{proximity-measure-part}
&& \lvert| g - f \rvert| <\epsilon \, , \\
\label{proximity-measure-antipart}
&& \lvert| \bar{g} - \bar{f} \rvert| < \bar{\epsilon} 
\end{eqnarray}
hold with arbitrary $\epsilon, \bar{\epsilon} >0$ for an appropriate
proximity measure. Since quasi-particles interact via their radiation
fields retardation constraints will be encountered in space-time.

\section{The radiation field}
Next, (\ref{maxwell-constraint-new}) is solved with the help of (\ref{quasi-part}) and
(\ref{quasi-antipart}). We obtain for the $4$-current
\begin{eqnarray}
\label{covariant-four-current}
&&\tensor{j}{^{\nu}} \left( \tensor{x}{^{\alpha}} \right) = 
-e \, \int \frac{d^3p}{p^0} \, c 
   \tensor{p}{^{\nu}} \, \left[ g \left( x, \vec{p} \right)- \bar{g}
   \left( x, \vec{p} \right) \right] \\
&&\hphantom{\tensor{j}{^{\nu}} \left( \tensor{x}{^{\alpha}} \right)}
   =-ec \sum_q \int d\tau_\mathrm{q} \,
   \left[ \tensor{u}{^{\nu}}_\mathrm{q}(\tau_\mathrm{q}) \, \delta^4 \left(
   \tensor{x}{^{\alpha}} - \tensor{x}{^{\alpha}}_\mathrm{q}(\tau_\mathrm{q}) \right) 
   \right. \nonumber \\
&&\hphantom{\tensor{j}{^{\nu}} \left( \tensor{x}{^{\alpha}} \right) =} \left. \hspace{3cm} 
   -\tensor{\bar{u}}{^{\nu}}_\mathrm{q}(\tau_\mathrm{q}) \, \delta^4 \left(
   \tensor{x}{^{\alpha}} - \tensor{\bar{x}}{^{\alpha}}_\mathrm{q}(\tau_\mathrm{q}) \right) 
   \right] \, . \nonumber 
\end{eqnarray}
We pick the retarded vector potential solutions of Maxwells's equations implying 
\begin{eqnarray}
\label{vector-potential}
&&\tensor{A}{^\mu}_\mathrm{ret}(\tensor{x}{^\alpha})= \frac{1}{\epsilon_0 c^2} \, \int d^4y \,
   G_\mathrm{ret} \left( \tensor{x}{^\alpha} -\tensor{y}{^\alpha}
   \right) \, \tensor{j}{^\mu} \left( \tensor{y}{^\alpha} \right) \, , 
\end{eqnarray}
where the retarded Green's function is given by
\begin{eqnarray}
\label{greens-function}
&& G_\mathrm{ret} \left( \tensor{z}{^\alpha} \right) = \frac{1}{2\pi} \, \Theta \left( z^0 \right) \, \delta 
   \left( z^2\right) \, , \quad \tensor{z}{^\alpha}=
   \tensor{x}{^\alpha}-\tensor{y}{^\alpha} \, .
\end{eqnarray}
Plugging (\ref{covariant-four-current}) and (\ref{greens-function})
into (\ref{vector-potential}) yields
\begin{eqnarray}
\label{vector-potential-explicit}
&&\hspace{-1.5cm}\tensor{A}{^\mu}_\mathrm{ret} \left( \tensor{x}{^\alpha} \right)
= -\frac{e}{\epsilon_0 c} \, \sum_q \int d\tau_\mathrm{q} \, \left[
   \tensor{u}{^\mu}_\mathrm{q}(\tau_\mathrm{q}) \, G_\mathrm{ret}
   \left( \tensor{x}{^\alpha} -
   \tensor{x}{^\alpha}_\mathrm{q}(\tau_\mathrm{q}) \right) \right. \\
&& \hspace{4cm} \left. -\tensor{\bar{u}}{^\mu}_\mathrm{q}(\tau_\mathrm{q}) \,
   G_\mathrm{ret} \left( \tensor{x}{^\alpha} -
   \tensor{\bar{x}}{^\alpha}_\mathrm{q}(\tau_\mathrm{q})
   \right) \right] \, . \nonumber
\end{eqnarray}
Defining $\tau_\mathrm{q \, ret}$ and
$\bar{\tau}_\mathrm{q \, ret}$ for particles and
anti-particles by
\begin{eqnarray}
\label{retardation-part}
&&\left[ x - x_\mathrm{q} \left( \tau_\mathrm{q \, ret}
   \right) \right]^2 =0 \, , \quad \tensor{x}{^0}_\mathrm{q} \left(
   \tau_\mathrm{q \, ret}
   \right) < \tensor{x}{^0} \, , \\
\label{retardation-antipart}
&&\left[ x - \bar{x}_\mathrm{q} \left( \bar{\tau}_\mathrm{q \, ret}
   \right) \right]^2 =0 \, , \quad \tensor{\bar{x}}{^0}_\mathrm{q}
   \left( \bar{\tau}_\mathrm{q \, ret}
   \right) < \tensor{x}{^0} 
\end{eqnarray}
we can solve (\ref{vector-potential}) by observing \cite{Itzykson1980}
\begin{eqnarray}
&&\hspace{-0.5cm}\Theta \left( \tensor{x}{^0} -
   \tensor{x}{^0}_\mathrm{q} (\tau_\mathrm{q}) \right) \, \delta \left[ \left( x - x_\mathrm{q}(\tau_\mathrm{q}) 
   \right)^2 \right] =\frac{1}{2 \left[ x - x_\mathrm{q}
   (\tau_\mathrm{q \, ret}
   ) \right] \cdot u_\mathrm{q} \left( \tau_\mathrm{q \, ret}
   \right)} \, , \\
&&\hspace{-0.5cm}\Theta \left( \tensor{x}{^0} -
   \tensor{\bar{x}}{^0}_\mathrm{q}(\tau_\mathrm{q}) \right) \, \delta \left[ \left( x - \bar{x}_\mathrm{q}(\tau_\mathrm{q}) 
   \right)^2 \right] =\frac{1}{2 \left[ x - \bar{x}_\mathrm{q}
   (\bar{\tau}_\mathrm{q \, ret}
   ) \right] \cdot \bar{u}_\mathrm{q} \left( \bar{\tau}_\mathrm{q \, ret}
   \right)} \, ,
\end{eqnarray}
since the particle worldlines intersect the backward light cone at the
observation point $\tensor{x}{^\alpha}$ for the retarded times. Hence,
we obtain the familiar retarded field solutions \cite{Itzykson1980}
\begin{eqnarray}
\label{retarded-vector-potential}
&&\tensor{A}{^\mu}_\mathrm{ret} \left( \tensor{x}{^\alpha} \right) =-\frac{e}{4\pi \epsilon_0
   c} \, \sum_q 
   \left[ \frac{\tensor{u}{^{\mu}}_\mathrm{q}\left(
   \tau_\mathrm{q \, ret} \right)}{ u_\mathrm{q} \left( \tau_\mathrm{q
   \, ret} \right) 
   \cdot \left[ x - x_\mathrm{q} \left( \tau_\mathrm{q\, ret}
   \right) \right]} \right. \\
&&\left. \hspace{5cm} - \frac{\tensor{\bar{u}}{^{\mu}}_\mathrm{q} \left(
   \bar{\tau}_\mathrm{q \, ret}
   \right)}{\bar{u}_\mathrm{q} \left( \bar{\tau}_\mathrm{q \, ret} \right) 
   \cdot \left[ x - \bar{x}_\mathrm{q} \left( \bar{\tau}_\mathrm{q \, ret}
   \right) \right]} \right] \, , \nonumber
\end{eqnarray}
where $\tensor{x}{^{\alpha}}_\mathrm{q} \left( \tau_\mathrm{q \, ret} \right)$ is the 
location of particle $q$ at its retarded time
$\tau_\mathrm{q \, ret} $. The same holds for the anti-particles
labeled with a bar. 

The radiation field (\ref{retarded-vector-potential}) is linked to the
worldlines of the particles and anti-particles. Hence, it is not defined for
all space-time points due to the retardation conditions
(\ref{retardation-part}) and (\ref{retardation-antipart}). 

\section{Equations of motion for quasi-particles}
To derive equations of motion for the quasi-particles we make use of
the energy-momentum tensors for the matter
(\ref{energy-momentum-spinor}) and radiation fields
(\ref{maxwell-constraint}). They are given by
\begin{eqnarray}
\label{energy-momentum-tensor-matter}
&&\tensor{t}{^{\mu}^{\nu}}^M = \int \frac{d^3p}{p^0} \,
   c \tensor{p}{^{\mu}} \tensor{p}{^{\nu}} \, \left( g + \bar{g}
   \right) 
\end{eqnarray}
and (\ref{energy-momentum-tensor-radiation}). To obtain an equation of
motion for (\ref{energy-momentum-tensor-matter}) along worldlines of
quasi-particles we make use of (\ref{vlasov-three-notation}). 

We pick an 
arbitrary quasi-particle at $\vec x_\mathrm{p} \left( t \right)$ and define a spherical
volume $V_\mathrm{p} \left( t \right)$ with radius $R_\mathrm{sp}
\left( t \right)$ surrounding it in such a way that there is no 2nd
quasi-particle at $\vec x_q \left( t \right)$ with $q \ne p$ in the
same volume. 

From (\ref{retarded-vector-potential})
we conclude that $V_\mathrm{p} \left( t \right)$ contains
the retarded field from the quasi-particle at $\vec{x}_\mathrm{p} \left(
t \right)$ inside  $V_\mathrm{p} \left( t \right)$ and the fields from
all quasi-particles at $\vec x_q \left( t \right)$ with $q \ne p$
outside $V_\mathrm{p} \left( t \right)$. The latter form the external field 
seen in  $V_\mathrm{p} \left( t \right)$ by quasi-particle $\vec{x}_\mathrm{p} \left(
t \right)$.

To shorten notation it is useful to split the total radiation field
into the source field $\tensor{F}{^{\mu}^{\nu}}_\mathrm{p}$ of
quasi-particle $p$ and the external field
$\tensor{F}{^{\mu}^{\nu}}_\mathrm{ext}$ produced by all quasi-particles $q \ne
p$. We obtain
\begin{eqnarray}
\label{eqn-of-motion-mechanics}
&&\partial_{\nu} 
\tensor{t}{^{\mu}^{\nu}}^M_\mathrm{p} = \left(
   \tensor{F}{^{\mu}^{\nu}}_\mathrm{p} +
   \tensor{F}{^{\mu}^{\nu}}_\mathrm{ext} \right) \,
   \tensor{j}{_{\nu}}_\mathrm{p} \, , \quad
   \tensor{F}{^\mu^\nu}_\mathrm{ext} =\sum_{q \ne p}
   \tensor{F}{^\mu^\nu}_\mathrm{q} \, .
\end{eqnarray}
To obtain an equation of motion for
(\ref{energy-momentum-tensor-radiation})
we consider only the field of quasi-particle $p$ inside the volume
$V_\mathrm{p}$. Hence, we find
\begin{eqnarray}
\label{eqn-of-motion-radiation}
&&\partial_{\nu} 
\tensor{t}{^{\mu}^{\nu}}^R_\mathrm{p} = - \tensor{F}{^{\mu}^{\nu}}_\mathrm{p} \,
   \tensor{j}{_{\nu}}_\mathrm{p} \, .
\end{eqnarray}
We note that only the field of quasi-particle $p$ contributes.
Adding  (\ref{eqn-of-motion-mechanics}) and
(\ref{eqn-of-motion-radiation}) we find
\begin{eqnarray}
\label{energy-momentum-tensor-total}
&&\partial_{\nu} \left( \tensor{t}{^\mu^\nu}^M_\mathrm{p} +
   \tensor{t}{^\mu^\nu}^R_p \right) = 
   \tensor{F}{^{\mu}^{\nu}}_\mathrm{ext} \,
   \tensor{j}{_{\nu}}_\mathrm{p} \, . 
\end{eqnarray}
Equation (\ref{energy-momentum-tensor-total}) does not contain
singular terms and can be used to define a set of delay equations 
for radiation reaction. We will not do this here but follow
the tradiational derivation of radiation reaction terms, which lead
us to the LAD equations.

\section{LAD equations}
We now solve (\ref{eqn-of-motion-radiation}) instead of
(\ref{energy-momentum-tensor-total}) explicitly for the
retarded field solution (\ref{retarded-vector-potential}). To do this
we first infer the current for source $p$ from (\ref{covariant-four-current})
and integrate over the volume $V_\mathrm{p}$ around $p$. We obtain
\begin{eqnarray}
\label{mean-field-eqn-of-motion-radiation}
&&\int_{V_\mathrm{p}} d^3x \left[ \partial_{\nu} 
\tensor{t}{^{\mu}^{\nu}}^R_\mathrm{p} \right] =
   \frac{q_\mathrm{p}}{\gamma(\tau_\mathrm{p})}
   \tensor{F}{^{\mu}^{\nu}}_\mathrm{p \, ret}
   \left[ x \left( \tau_\mathrm{p} \right)  \right] \, 
   \tensor{u}{_{\nu}}_\mathrm{p} \left( \tau_\mathrm{p} \right) \, .
\end{eqnarray}
We next evaluate 
\begin{eqnarray}
&&\tensor{F}{^{\mu}^{\nu}}_\mathrm{p \, ret}
   \left[ \tensor{x}{^\alpha}  \right] \, 
   \tensor{u}{_{\nu}}_\mathrm{p}(\tau_\mathrm{p}) 
\end{eqnarray}
following the outline in \cite{Hartemann200112}. It is found after a
few intermediate steps
\begin{eqnarray}
\label{mean-field-eqn-of-motion-radiation1}
&&\int_{V_\mathrm{p}} d^3x \left[ \partial_{\nu} 
\tensor{t}{^{\mu}^{\nu}}^M_\mathrm{p} \right] +\left( \frac{\tau_0}{2 \gamma(\tau_\mathrm{p})}
   \int^{\infty}_0 ds \, \frac{\delta \left( s \right)}{s} \right) 
   \tensor{a}{^\mu}_\mathrm{p} \left( \tau_\mathrm{p} \right) \\
&&= \frac{q_\mathrm{p}}{\gamma(\tau_\mathrm{p})} \tensor{F}{^{\mu}^{\nu}}_\mathrm{ext}
   \left[ x \left( \tau_\mathrm{p} \right)  \right] \, 
   \tensor{u}{_{\nu}}_\mathrm{p} \left( \tau_\mathrm{p} \right) 
  + \frac{2\tau_0}{3 \gamma(\tau_\mathrm{p})} \left(
   \tensor{\dot{a}}{^\mu}_\mathrm{p}
  +\frac{a_\mathrm{p} \cdot a_\mathrm{p}}{c^2} \,
   \tensor{u}{^\mu}_\mathrm{p} \right) \left( \tau_\mathrm{p} \right) 
   \, , \nonumber
\end{eqnarray} 
where $\tau_0 = q^2_\mathrm{p}/4\pi \epsilon_0 c^3$ and
$\tensor{a}{^\mu}_\mathrm{p}$ is the 4-acceleration. Finally, we
obtain the following set of ordinary differential equations for each
quasi-element $p$
\begin{eqnarray}
\label{mean-field-eqn-of-motion-radiation2a}
&&\tensor{\dot{x}}{^\mu}_\mathrm{p}\left( \tau_\mathrm{p} \right) =
   \tensor{u}{^\mu}_\mathrm{p} \left( \tau_\mathrm{p} \right) \, ,\\
\label{mean-field-eqn-of-motion-radiation2b}
&&\left( m_\mathrm{p} +\frac{\tau_0}{2}
   \int^{\infty}_0 ds \, \frac{\delta \left( s \right)}{s} \right) 
   \tensor{\dot{u}}{^\mu}_\mathrm{p} \left( \tau_\mathrm{p} \right) \\
&&= q_\mathrm{p} \left[ \tensor{F}{^{\mu}^{\nu}}_\mathrm{ext} \, 
   \tensor{u}{_{\nu}}_\mathrm{p} \right] \left( \tau_\mathrm{p} \right) 
+ \frac{2\tau_0}{3} \left( \tensor{\ddot{u}}{^\mu}_\mathrm{p}
  +\frac{a_\mathrm{p} \cdot a_\mathrm{p}}{c^2} \,
   \tensor{u}{^\mu}_\mathrm{p} \right) \left( \tau_\mathrm{p} \right) \, , \nonumber 
\end{eqnarray}
the solutions of which have to be plugged into
(\ref{retarded-vector-potential}) to obtain the field distribution of
the particle and anti-particle ensemble. 

The self-force terms in (\ref{mean-field-eqn-of-motion-radiation2b})
are part of the LAD equations, which have well-known mathematical
problems \cite{Hartemann200112}. We note that the derivation of the LAD
equations given here makes the assuption that the radiation fields 
can be Taylor expanded around their singularities along the 
worldlines. No need for similar Taylor expansions would arise in the case
of the aforementioned delay equations as a replacement for LAD.

\section{The dynamical framework}
\begin{figure}[h]
\begin{center}
\includegraphics[width=0.5\linewidth]{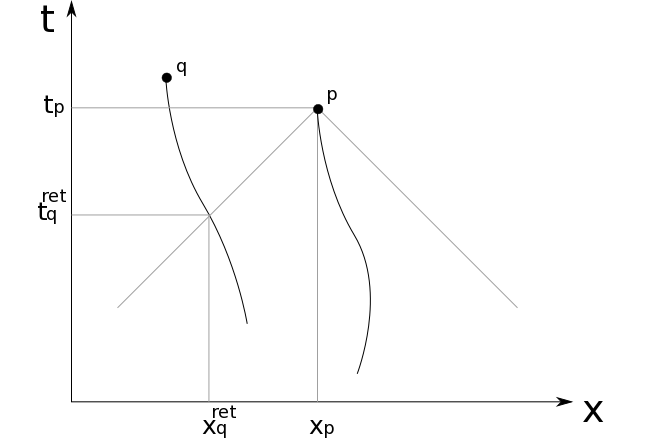}
\end{center}
\caption[minkowski space-time]{\label{fig_spacetime}
 Intersection of the backward light cone of the observation point at
 $x^\mu_p$ with the worldline $x^\mu_q(\tau)$ of particle $q$
 in Minkowski space. }
\end{figure}

Taking all together we obtain a set of classical MD 
equations of motion given by
\begin{eqnarray}
\label{energy-conservation3a}
&&\tensor{\dot{x}}{^\alpha}_\mathrm{p} =
   \tensor{u}{^\alpha}_\mathrm{p} \, ,\\
\label{mean-field-eqn-of-motion-radiation3b}
&&\bar{m}_\mathrm{p} \,  
   \tensor{\dot{u}}{^\alpha}_\mathrm{p} = q_\mathrm{p} \tensor{F}{^{\alpha}^{\nu}}_\mathrm{ext} \, 
   \tensor{u}{_{\nu}}_\mathrm{p}  + \frac{2\tau_0}{3} \left( \tensor{\ddot{u}}{^\alpha}_\mathrm{p}
  +\frac{a_\mathrm{p} \cdot a_\mathrm{p}}{c^2} \,
   \tensor{u}{^\alpha}_\mathrm{p} \right) \, , \nonumber 
\end{eqnarray}
where $\bar{m}_\mathrm{p}$ denotes the renormalized mass
\begin{eqnarray}
&&\bar{m}_\mathrm{p}=m_\mathrm{p} +\frac{\tau_0}{2}
   \int^{\infty}_0 ds \, \frac{\delta \left( s \right)}{s} \, .
\end{eqnarray}
The external electromagnetic field at $\tensor{x}{^\alpha}_\mathrm{p}$
is generated by all surrounding particles with
$\tensor{x}{^\alpha}_\mathrm{p} \ne \tensor{x}{^\alpha}_\mathrm{q}$
\begin{eqnarray}
\label{energy-conservation3d}
&&\tensor{F}{^\mu^{\nu}}_\mathrm{ext} \left(
   \tensor{x}{^\alpha}_\mathrm{p} \right) = \sum_{\mathrm{q} \ne
   \mathrm{p}} \tensor{F}{^\mu^{\nu}}_\mathrm{q} \left(
   {x}{^\alpha}_\mathrm{p} \right) \, , \\
\label{energy-conservation3e}
&&\tensor{F}{^\mu^{\nu}}_\mathrm{q} \left(
   {x}{^\alpha}_\mathrm{p} \right) = \left(
\begin{array}{cccc}
0 & -\frac{\tensor{E}{^1}_\mathrm{q}}{c} & -\frac{\tensor{E}{^2}_\mathrm{q}}{c} & -\frac{\tensor{E}{^3}_\mathrm{q}}{c} \\ 
\frac{\tensor{E}{^1}_\mathrm{q}}{c} & 0 & -\tensor{B}{^3}_\mathrm{q} & \tensor{B}{^2}_\mathrm{q} \\ 
\frac{\tensor{E}{^2}_\mathrm{q}}{c} & \tensor{B}{^3}_\mathrm{q} & 0 & -\tensor{B}{^1}_\mathrm{q} \\ 
\frac{\tensor{E}{^3}_\mathrm{q}}{c} & -\tensor{B}{^2}_\mathrm{q} & \tensor{B}{^1}_\mathrm{q} & 0 
\end{array}
\right) \left(
   {x}{^\alpha}_\mathrm{p} \right) \, .
\end{eqnarray}
The retarded electromagnetic fields are obtained from
(\ref{retarded-vector-potential}). They are given by \cite{Itzykson1980}
\begin{eqnarray}
\label{energy-conservation3f}
&&\tensor{E}{^i}_\mathrm{q} \left( {x}{^\alpha}_\mathrm{p} \right) =
                                                         \frac{e}{4\pi\epsilon_0}
                                                         \left(
                                                         \frac{ c^2 \,
   \left( \tensor{n}{^i}_\mathrm{q}-\tensor{\beta}{^i}_\mathrm{q}
   \right)}{ \left( {u}{^0}_\mathrm{q} \right)^2 
                                                         \, (1-\vec
                                                         \beta_\mathrm{q}
                                                         \cdot \vec
                                                         n_\mathrm{q})^3
                                                         R^2_\mathrm{q}}
                                                         \vphantom{\frac{ \tensor{\left[ \vec n_\mathrm{q} \times \left( (\vec n_\mathrm{q}-\vec
    \beta_\mathrm{q}) \times 
    \dot{\vec \beta}_\mathrm{q} \right)  \right]}{^i}}{c \, \left( 1-\vec
  \beta_\mathrm{q} \cdot \vec n_\mathrm{q} \right)^3 R_\mathrm{q}}}
\right. \\
&&\hphantom{\tensor{E}{^i}_\mathrm{q} \left( {x}{^\alpha}_\mathrm{p} \right) =}
\left. + \frac{ \tensor{\left[ \vec n_\mathrm{q} \times \left( (\vec n_\mathrm{q}-\vec
    \beta_\mathrm{q}) \times 
    \dot{\vec \beta}_\mathrm{q} \right)  \right]}{^i}}{{u}{^0}_\mathrm{q} \, \left( 1-\vec
  \beta_\mathrm{q} \cdot \vec n_\mathrm{q} \right)^3 R_\mathrm{q}}
 \right) \left(
   {x}{^\alpha}_\mathrm{p} \right) \, , \nonumber \\
\label{energy-conservation3g}
&&\tensor{B}{^i}_\mathrm{q} \left( {x}{^\alpha}_\mathrm{p} \right)  =
   \frac{1}{c} \,
                                   \tensor{\epsilon}{^{i}^{j}^{k}} \,
   \tensor{n}{^j}_\mathrm{q} \,
                          \tensor{E}{^k}_\mathrm{q} \left(
   {x}{^\alpha}_\mathrm{p} \right)  \, ,
\end{eqnarray}
where
\begin{eqnarray}
\label{energy-conservation3h}
&&R_\mathrm{q} = |\vec{{x}} - \vec{{x}}^\mathrm{\; ret}_\mathrm{q}| \, , \quad
\tensor{n}{^i}_\mathrm{q}=\frac{\tensor{{x}}{^i} -
   \tensor{{x}}{^i}^\mathrm{\; ret}_\mathrm{q}}{R_\mathrm{q}} \, , \\
&&\tensor{{\beta}}{^i}_\mathrm{q}=\frac{\tensor{\dot{{x}}}{^i}^\mathrm{\;
   ret}_\mathrm{q}}{c} \, , \quad  u^0_\mathrm{q}=\dot{{x}}{^0}^\mathrm{\;
   ret}_\mathrm{q} \, . 
\end{eqnarray}
The retardation constraint is
\begin{align}
\label{t_ret}
 c \, \left( {t}  - {t}^\mathrm{\; ret}_\mathrm{q} 
  \right) &= \left| \vec{{x}} - \vec{{x}}^\mathrm{\;
            ret}_\mathrm{q} \right| \, ,
\end{align}
where
\begin{align}
\label{x_ret}
\tensor{x}{^\mu}^\mathrm{\;
  ret}_\mathrm{q}&=\tensor{x}{^\mu}_\mathrm{q} \left( \tau_\mathrm{q
                   \, ret} \right) \, .
\end{align}
Constraint (\ref{t_ret}) must be solved for all $\mathrm{q}
\ne \mathrm{p}$. In case a solution exists particle $\mathrm{q}$
contributes to the external field at particle $\mathrm{p}$. Else it
does not. The situation is illustrated in Fig. \ref{fig_spacetime}.
The discussion of the setup problem of the radiative MD system is
omitted in this paper.

\section{Conclusions} \label{conclusions}
Starting from the field equations of electrodynamics equations of
motion for scalar quasi-electrons and positrons have
been derived. Together with their radiation fields they form a set of
MD equations with self-force effects. The derivation of the latter
makes use of the energy-momentum tensors of the matter and 
radiation fields. In the paper the LAD terms for radiation reaction
have been motivated.



  \bibliographystyle{elsarticle-num} 
  \bibliography{literatur_eqn_motion}





\end{document}